\documentclass[12pt]{iopart}

\usepackage{epsfig,latexsym,amsmath}
\usepackage[abbr]{harvard} 

\newcommand{\HALF}{{\textstyle\frac{1}{2}}}

\newcommand{\SHALF}{{\scriptstyle\frac{1}{2}}}

\newcommand{\SFRAC}[2]{{\scriptstyle\frac{#1}{#2}}}

\newcommand{\EMB}{\boldsymbol}
\newcommand{\VEC}[1]{{\EMB{#1}}}
\newcommand{\LAB}[1]{{\mathsf{#1}}}
\newcommand{\FUN}[1]{{\mathrm{#1}}}

\newcommand{\TR}{\FUN{tr}}

\newcommand{\UPS}{\EMB\iota}
\newcommand{\IX}{\VEC I_\LAB{x}}
\newcommand{\IY}{\VEC I_\LAB{y}}
\newcommand{\IZ}{\VEC I_\LAB{z}}

\newcommand{\EP}{\VEC E_{+}}
\newcommand{\EM}{\VEC E_{-}}

\newcommand{\KET}[1]{|\,#1\,\rangle}

\newcommand{\AVG}[1]{\langle\,#1\,\rangle}

\newcommand{\spinA}{{\LAB{A}}}
\newcommand{\spinB}{{\LAB{B}}}
\newcommand{\axisX}{{\LAB{x}}}

\newcommand{\axisZ}{{\LAB{z}}}
\newcommand{\plus}{{+}}
\newcommand{\minus}{{-}}

\newcommand{\Heff}{\VEC H_{\mathrm{eff}}}
\newcommand{\Htrn}{\VEC H_{\mathrm{trn}}}

\begin{document}

\title{The effective Hamiltonian of the
Pound-Overhauser controlled-NOT gate}
\author{David G Cory\dag, Amy E Dunlop\dag,
Timothy F Havel\ddag\footnote[3]{To whom correspondence
may be e-mailed at \texttt{tfhavel@mit.edu}.},
Shyamal S Somaroo\ddag{} and Wurong Zhang$\|$}
\address{\dag\ Department of Nuclear Engineering, Massachusetts,
Institute of Technology, Cambridge, MA 02139, USA}
\address{\ddag\ Biological Chemistry and Molecular Pharmacology,
Harvard Medical School, Boston, MA 02115, USA}
\address{$\|$ Bruker Instruments, Billerica, MA 01821, USA}
\begin{abstract}
In NMR-based quantum computing, it is known that
the controlled-NOT gate can be implemented by applying
a low-power, monochromatic radio-frequency field to one
peak of a doublet in a weakly-coupled two-spin system.
This is known in NMR spectroscopy as Pound-Overhauser double resonance.
The ``transition'' Hamiltonian that has been associated
with this procedure is however only an approximation,
which ignores off-resonance effects and does not
correctly predict the associated phase factors.
In this paper, the exact effective Hamiltonian for
evolution of the spins' state in a rotating frame is derived,
both under irradiation of a single peak (on-transition)
as well as between the peaks of the doublet (on-resonance).
The accuracy of these effective Hamiltonians is
validated by comparing the observable product operator
components of the density matrix obtained by simulation
to those obtained by fitting the corresponding experiments.
It is further shown that an on-resonance field yields a new
implementation of the controlled-NOT gate up to phase factors,
wherein the field converts the $\IZ^\spinA$ state
into the antiphase state $2 \IX^\spinA \IZ^\spinB$,
which is then converted into the desired two-spin
order $2 \IZ^\spinA \IZ^\spinB$ by a broadband
$\pi/2$ pulse selective for the $\spinA$ spin.
In the on-transition case, it is explained
that while a controlled-NOT gate is approximately
obtained whenever the radio-frequency field
power is low compared to the spin-spin coupling,
at certain specific power levels an exact
implementation is obtained up to phase factors.
For both these implementations, the phase factors
are derived exactly, enabling them to be corrected.
In Appendices, the on-resonance Hamiltonian
is analytically diagonalized, and proofs are
given that, in the weak-coupling approximation,
off-resonance effects can be neglected whenever the
radio-frequency field power is small compared to the
difference in resonance frequencies of the two spins.
\end{abstract}
\jl{1} \pacs{05.30.-d, 75.45.+j, 76.70.Fz, 89.80.+h}
\date{\today} \maketitle
\section{INTRODUCTION}
The c-NOT (controlled-NOT) gate is of central importance
in quantum computing \cite{BarDeuEke:95,BBCDMSSSW:95}.
It is well-known \cite{Lloyd:93}
that in quantum computers based on
frequency addressing of their q-bits,
such as solution-state NMR spectroscopy on spin
$\HALF$ nuclei \cite{CorFahHav:97,GershChuan:97},
the c-NOT gate can be implemented by the
application of a two q-bit transition
Hamiltonian of the form\footnote[1]{
Throughout this paper we shall be making use of
the \emph{product operator} notation that is
widely used in NMR spectroscopy \cite{SoEiLeBoEr:83},
and the rules for manipulating these symbols derived
from geometric algebra \cite{SomCorHav:98}.}
\begin{equation}
\Htrn ~\equiv~
\pi \, \IX^\spinA \EM^\spinB
\equiv~ \tfrac{\pi}{2}
\begin{pmatrix} 0 & 0 & 0 & 0 \\ 0 & 0 & 0 & 1 \\
0 & 0 & 0 & 0 \\ 0 & 1 & 0 & 0 \end{pmatrix} ~,
\end{equation}
where $\VEC E_{\pm}^\spinB \equiv \HALF
(1 {\pm} 2 \IZ^\spinB)$ is idempotent
(cf.\ \cite{HatanYanno:81}).
Using NMR spectroscopy \cite{CorPriHav:98},
we have previously demonstrated that irradiating exactly one peak of
a doublet in a two-spin system with a power $\omega_1$ much less than
the coupling between the spins $2\pi J^{\spinA\spinB}$ transforms
the equilibrium state in accord with the corresponding propagator,
i.e.\ $\exp(-\UPS t \Htrn)$
\begin{equation}
~=~ e^{-\UPS \pi t \IX^\spinA}
\EM^\spinB + \EP^\spinB
=~ \begin{pmatrix}
~1~ & 0 & 0 & 0 \\ 0 & \cos(\pi t / 2) & 0 & -\UPS \sin( \pi t / 2) \\
0 & 0 & ~1~ & 0 \\ 0 & -\UPS \sin( \pi t / 2) & 0 & \cos(\pi t / 2)
\end{pmatrix} ~.
\end{equation}
In NMR spectroscopy, this is often called
{\em Pound-Overhauser double resonance\/} \cite{Slichter:90}.
The nonzero phase factors in this matrix can be equalized
by a $\pi/2$ evolution $\exp(-\UPS (\pi/2) \IZ^\spinB)$.
Nevertheless, the application of this procedure to
a (mixture of) superposition states quickly shows
that the effective Hamiltonian is not so simple.
The transition Hamiltonian is only a nonphysical
approximation, which cannot predict all the details
of the spins' evolution under selective irradiation.

In this paper the effective Hamiltonian in a rotating
frame \cite{ErnBodWok:87,Slichter:90} is derived,
which fully describes the evolution of a weakly-coupled,
two-spin system under a monochromatic RF (radio-frequency)
field, both on-resonance as well as on a single transition.
A pictorial representation in terms of effective
fields is described, which provides an intuitive
description of the spin dynamics under these Hamiltonians.
The results of NMR experiments are presented, which
demonstrate that the superposition states that evolve
under monochromatic RF fields are consistent with
those obtained from simulations using these effective
Hamiltonians, and with the effective fields picture.
Assuming $\omega_1 \ll 2\pi |J^{\spinA\spinB}| \ll
|\omega^\spinA - \omega^\spinB|$ and that off-resonance
effects can be neglected, it is proven that the transition
and effective on-transition Hamiltonians are equivalent,
in the sense that the corresponding propagators
are approximately equal up to conditional phases
(i.e.\ a diagonal matrix of phase factors).
This implementation of the c-NOT is shown
to be exact (up to conditional phases) for
$\omega_1 = 2\pi J^{\spinA\spinB}/\sqrt{4n^2-1}$
where $n > 0$ is an integer.
It is further proven that the application of an on-resonance
field with $\omega_1 = \pi|J^{\spinA\spinB}|$ followed by a
broadband $\pi/2$ ``soft pulse'' (covering the entire doublet),
both on the $\spinA$ spin, likewise implements
the c-NOT gate up to conditional phases.
In both cases, the conditional phases are derived explicitly.
In appendices, the on-resonance Hamiltonian is analytically diagonalized,
and the assumption that off-resonance effects are negligible
whenever $\omega_1 \ll |\omega^\spinA - \omega^\spinB|$ is
rigorously justified under the weak-coupling approximation.

\section{THE POUND-OVERHAUSER EFFECTIVE HAMILTONIAN}

\subsection{Derivation of the effective Hamiltonians}
Given a Hamiltonian $\VEC H = \VEC H(t)$ and unitary
transformation $\VEC U \equiv \exp(-\UPS\VEC G t)$
(where $\VEC G = \tilde{\VEC G}$ is Hermitian),
the evolution of the transformed density matrix
$\EMB\rho' = \VEC U \EMB\rho\,\tilde{\VEC U}$
is given by \cite{ErnBodWok:87,Slichter:90}
\begin{equation} \label{eq:tfm} \begin{split}
\dot{\EMB\rho}' ~=~ &
\dot{\VEC U}\EMB \rho \, \tilde{\VEC U}
+ \VEC U \dot{\EMB \rho} \, \tilde{\VEC U}
+ \VEC U \EMB \rho \, \dot{\tilde{\VEC U}}
\\ =~ &
- \UPS \VEC G \EMB \rho'
+ \UPS [\EMB \rho', \VEC H']
+ \UPS \EMB \rho' \VEC G
\\ =~ &
\UPS [\EMB \rho', \VEC H' + \VEC G] ~\equiv~
\UPS [\EMB \rho', \Heff] ~.
\end{split} \end{equation}
where $\VEC H' \equiv \VEC U \VEC H \tilde{\VEC U}$
and $\Heff \equiv \VEC H' + \VEC G$.
The weak-coupling Hamiltonian of
a two-spin system is\footnote{
In quantum computing, the ground
state is generally indexed by ``$0$'',
whereas NMR spectroscopists typically
put the spin ``up'' (parallel the field)
state before the ``down'' in their matrices.
These two conventions agree only when
the gyromagnetic ratio is positive,
as will be assumed in this paper.}
\begin{equation} \label{eq:weakham}
\VEC H^{\spinA\spinB} ~=~
- \omega_0^\spinA \IZ^\spinA
- \omega_0^\spinB \IZ^\spinB
+ 2\pi J^{\spinA\spinB} \IZ^\spinA \IZ^\spinB ~,
\end{equation}
where $\omega^\spinA$, $\omega^\spinB$ are
the Larmour precession frequencies of the spins,
and $J^{\spinA\spinB}$ the scalar coupling between them.
The applied RF field Hamiltonian has the form
\begin{equation}
\VEC H_{\mathrm{RF}} ~=~ \omega_1
e^{\UPS t \VEC G} (\IX^\spinA + \IX^\spinB) e^{-\UPS t \VEC G} ~,
\end{equation}
where the RF power transmitted to the
spins is (assuming for convenience that
the system is homonuclear) $\omega_1 > 0$,
$\VEC G \equiv \omega_2(\IZ^\spinA+\IZ^\spinB)$,
and $\omega_2$ is the frequency of the RF field.

These results show that the
time-dependence can be removed from the Hamiltonian
$\VEC H = \VEC H^{\spinA\spinB} + \VEC H_{\mathrm{RF}}$
by transformation to a frame rotating at the frequency $\omega_2$.
If the frequency $\omega_2 = \omega_0^\spinA + \pi J^{\spinA\spinB}$
matches the $\KET{01} \leftrightarrow \KET{11}$
component of the $\spinA$-spin doublet, this yields
the time-independent ``effective Hamiltonian''
\begin{equation} \label{eq:ontrn}
\Heff ~=~ \pi J^{\spinA\spinB} (\IZ^\spinA + 2 \IZ^\spinA \IZ^\spinB)
+ (\omega_0^\spinA - \omega_0^\spinB + \pi J^{\spinA\spinB}) \IZ^\spinB
+ \omega_1 (\IX^\spinA + \IX^\spinB) ~.
\end{equation}
In the event that the RF field is placed on-resonance,
$\omega_2 = \omega_0^\spinA$, and hence
\begin{equation} \label{eq:onres}
\Heff ~=~ 2 \pi J^{\spinA\spinB} \IZ^\spinA \IZ^\spinB
+ (\omega_0^\spinA - \omega_0^\spinB) \IZ^\spinB
+ \omega_1 (\IX^\spinA + \IX^\spinB) ~.
\end{equation}

Note that since $\VEC G \equiv \omega_2(\IZ^\spinA + \IZ^\spinB)$
commutes with $\IX^\spinA\IX^\spinB + \IY^\spinA\IY^\spinB$,
if we use the full strong coupling Hamiltonian
\begin{equation}
\VEC H^{\spinA\spinB} ~=~ - \omega_0^\spinA \IZ^\spinA
- \omega_0^\spinB \IZ^\spinB + 2\pi J^{\spinA\spinB}
\left( \IX^\spinA\IX^\spinB + \IY^\spinA\IY^\spinB
+ \IZ^\spinA \IZ^\spinB \right) ~.
\end{equation}
instead of its weak coupling approximation (\ref{eq:weakham}),
Eqs.\ (\ref{eq:ontrn}) and (\ref{eq:onres}) need only be
modified by the addition of the term $2\pi J^{\spinA\spinB}
( \IX^\spinA\IX^\spinB + \IY^\spinA\IY^\spinB )$.

\begin{figure}[tb] \begin{center}
\psfig{file=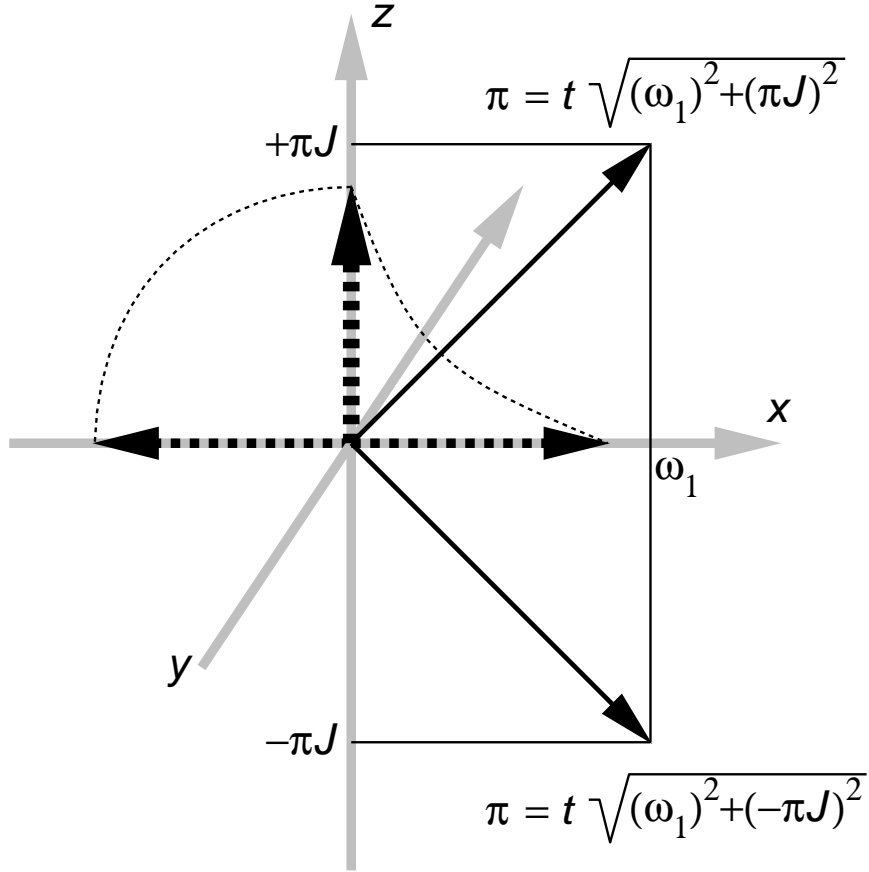,width=4.5in}
\end{center} \caption{
Effective fields picture of the evolution of the
irradiated $\spinA$-spins in the on-resonance case.
There are two subpopulations of molecules:
that in which the $\spinB$-spin is ``up'',
and that in which the $\spinB$-spin is ``down''.
In these two subpopulations, the magnetization
due to the $\spinA$-spins is initially aligned
with the applied field (broad banded up arrow).
The RF field power is $\omega_1 = \pi |J^{\spinA\spinB}|$,
which in the co-rotating frame yields the
effective fields shown with the thin solid
arrows inclined at $\pi/4$ from the $z$-axis.
After a time $t = \pi / ((\omega_1)^2
+ (\pi J^{\spinA\spinB})^2)^{-1/2}$,
these fields have rotated the $\spinA$-spin in both
subpopulations by an angle of $\pi$ to the $\pm x$-axis,
which corresponds to the antiphase state $\IX^\spinA\IZ^\spinB$.
} \label{fig:on-res} \end{figure}

\begin{figure}[tb] \begin{center}
\psfig{file=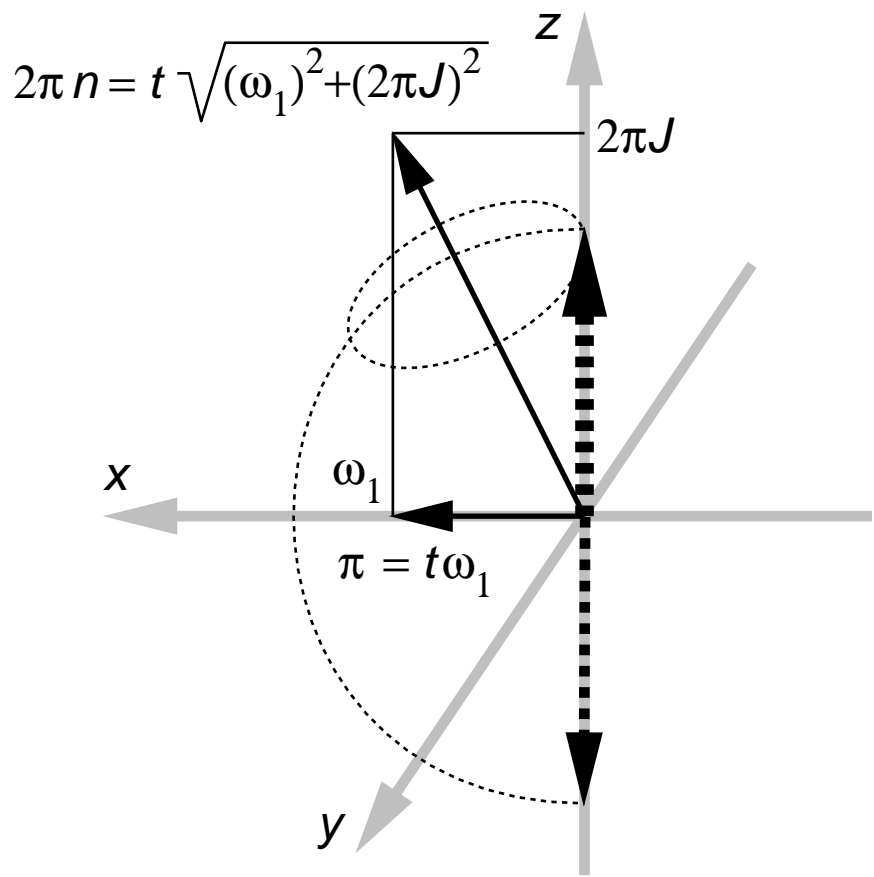,width=4.5in}
\end{center} \caption{
Effective fields picture of the evolution of the
irradiated $\spinA$-spins in the on-transition case.
As in Figure \ref{fig:on-res},
the magnetization due to the $\spinA$-spins in the
two subpopulations of molecules is initially aligned
with the applied field (broad banded up arrow).
The RF field power is $\omega_1 \ll \pi |J^{\spinA\spinB}|$,
which in the co-rotating frame yields the
effective fields shown with the thin solid arrows.
The one corresponding to the irradiated transition
is along the $x$-axis, while the other one is displaced
towards the $z$-axis by an amount $2\pi J^{\spinA\spinB}$.
After a time $t = \pi / \omega_1$,
the $\spinA$-spin in the first subpopulation
has been rotated by $\pi$ to the $-z$-axis.
If $\omega_1$ is set so that at this time
the $\spinA$-spin in the other subpopulation has
rotated by a multiple of $2\pi$, this yields an exact
implementation of the c-NOT up to conditional phases.
} \label{fig:on-trn} \end{figure}

\subsection{The effective fields picture of the evolution}
A physical picture of the spin dynamics may be obtained by
breaking the NMR spectrum up into its individual resonance
lines, and specifying a distinct effective field for each.
This picture assumes we are dealing with an equilibrium
density matrix in the high-temperature approximation,
so that there are no spin-spin correlations,
and neglects relaxation effects.
In a frame rotating at the transmitter frequency $\omega_2$,
the residual static magnetic field at each resonance $\omega_0$
is equal to the frequency offset $\omega_0 - \omega_2$.
Each of these residual fields may be identified
with a subpopulation of the molecules present,
wherein the other spins are aligned so that their
couplings with the spin in question cause it to
resonate at the frequency of the residual field.
The applied radio-frequency field contributes
a transverse component of strength $\omega_1$
to the net effective field for each resonance.
With care, all the results of this paper
can be derived by purely geometric means
from this ``effective fields picture''.

In the on-resonance case with
$\omega_1 = \pi |J^{\spinA\spinB}|$,
the effective fields for each resonance
line of the $\spinA$-spin doublet form
angles of $\pm\pi/4$ with the $z$-axis,
as shown in Figure \ref{fig:on-res}.
Thus the equilibrium magnetization components
for each resonance counter-rotate from their initial
positions along the $z$-axis to opposite directions
along the $x$-axis after a time ``$t$'' given by
$\pi = t \sqrt{(\omega_1)^2 + (\pi J^{\spinA\spinB})^2}$.
This vector configuration represents
the antiphase state $\IX^\spinA\IZ^\spinB$,
which may be converted to the $\IZ^\spinA\IZ^\spinB$
state expected after a c-NOT gate by a broadband
$\pi/2$ $y$-pulse selective for the $\spinA$-spin.
In the on-transition case, in contrast,
the on-resonance component of the magnetization
experiences an effective field in the transverse plane,
while the other component experiences a field which is,
for $\omega_1 \ll |\omega_0^\spinA - \omega_0^\spinB|$,
very nearly along the $z$-axis.
Since it is exactly along the $z$-axis only
in the limit $\omega_1 \rightarrow 0$, however,
for any $\omega_1 > 0$ it nutates away from the
$z$-axis, as shown in Figure \ref{fig:on-trn}.
An exact implementation of the c-NOT gate is
nevertheless obtained when $\omega_1$ is chosen so
that this component is rotated by an integer multiple
of $2\pi$ back to the $z$-axis in the time it takes
the other component to rotate to the $-z$-axis,
namely $t = \pi / \omega_1$.

\section{EXPERIMENTAL VALIDATION OF THE EFFECTIVE HAMILTONIAN}
\begin{figure}[tb] \begin{center}
\epsfig{file=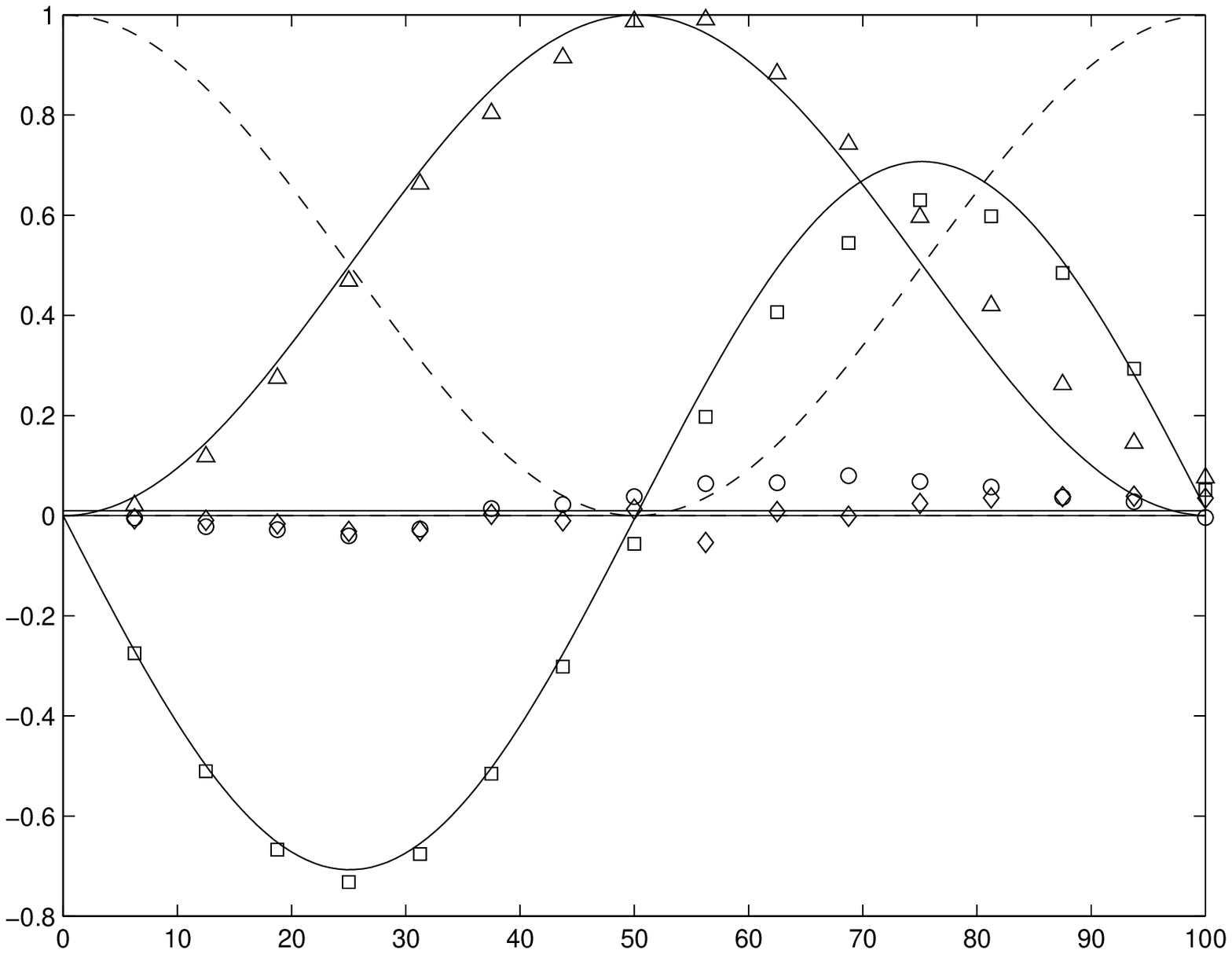,width=5.0in}
\end{center} \caption{
Plots of observable product operator components
of the density matrix versus 16 equally spaced
time points (in percent of $t_{\mathrm{max}}$)
for the on-resonance case \textbf{(i)}
with $\omega_1 = \pi |J^{\spinA\spinB}|$ and
$t_{\mathrm{max}} = \sqrt{2} / |J^{\spinA\spinB}|$.
The symbols in this plot (and the
corresponding product operator components) are
``$\scriptstyle\bigcirc$'' ($\IX^\spinA$),
``$\Box$'' ($\IY^\spinA$),
``$\bigtriangleup$'' ($\IX^\spinA \IZ^\spinB$), and
``{\Large$\diamond$}'' ($\IY^\spinA \IZ^\spinB$).
The solid lines are the simulated results for these components,
while the dashed line is the simulated evolution of
the diagonal $\IZ^\spinA$ component.
} \label{fig:seriesA} \end{figure}

\begin{figure}[tb] \begin{center}
\epsfig{file=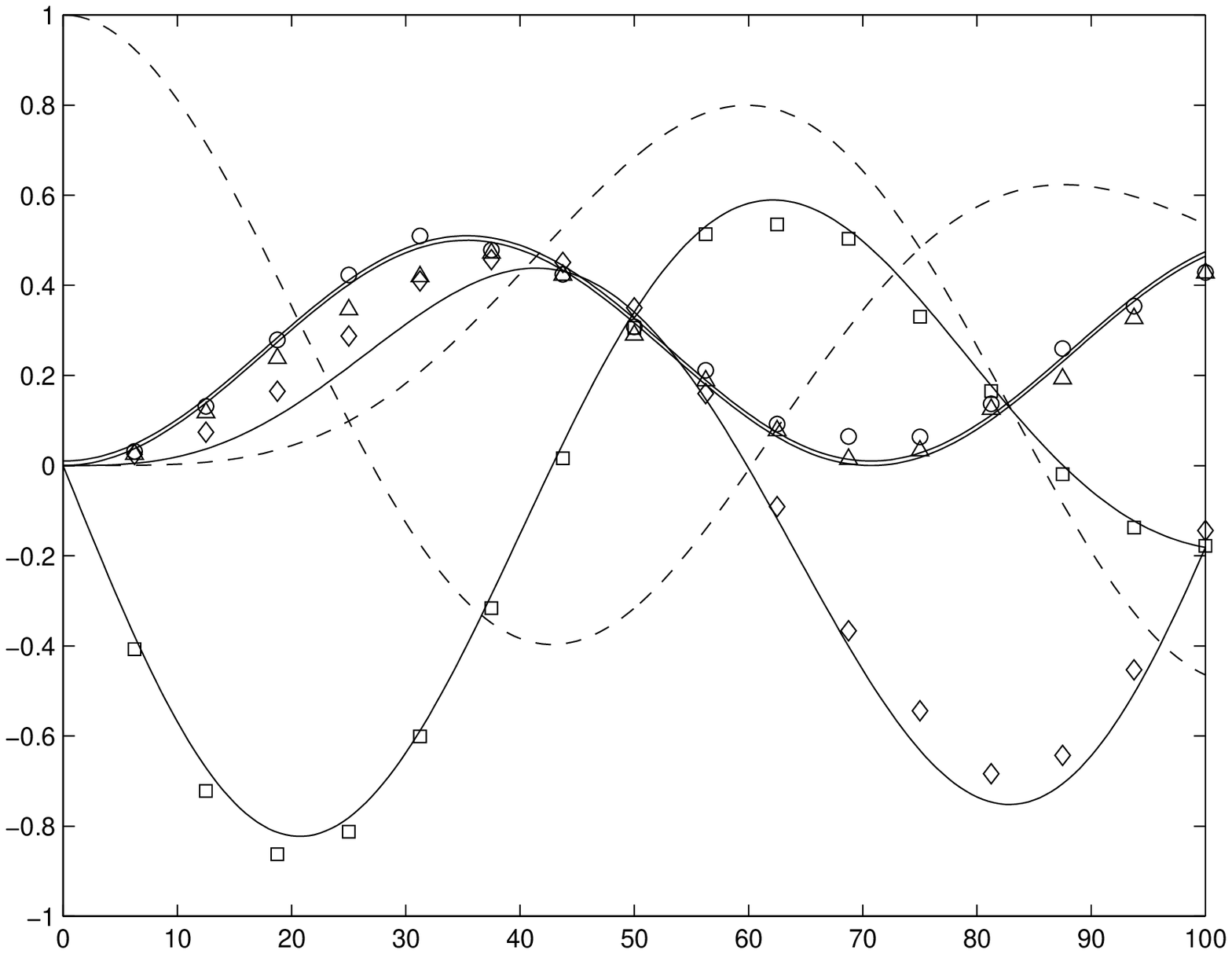,width=5.0in}
\end{center} \caption{
Plots of observable product operator components
of the density matrix versus 16 equally spaced
time points (in percent of $t_{\mathrm{max}}$)
for the on-transition case \textbf{(ii)}
with $\omega_1 = 2 \pi |J^{\spinA\spinB}|$ and
$t_{\mathrm{max}} = 1 / |J^{\spinA\spinB}|$.
The solid lines are the simulated results for these
components, while the dashed line depicts the
evolution of the diagonal $\IZ^\spinA$
(starting from $1$) and $\IZ^\spinA
\IZ^\spinB$ (starting from $0$) components.
The symbols in the plot are the same
as in Figure \ref{fig:seriesA}.
} \label{fig:seriesB} \end{figure}

\begin{figure}[tb] \begin{center}
\epsfig{file=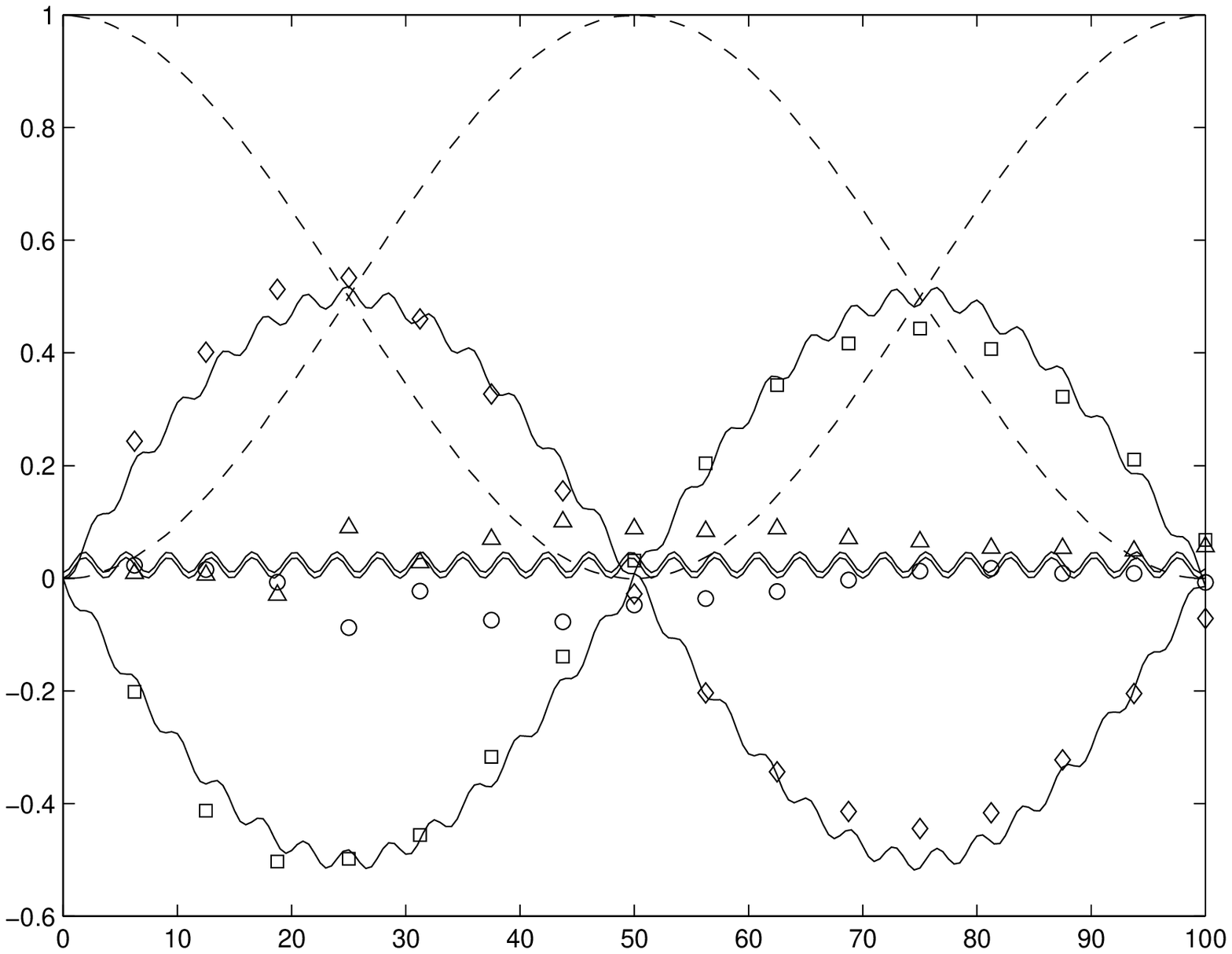,width=5.0in}
\end{center} \caption{
Plots of observable product operator components
of the density matrix versus 16 equally spaced
time points (in percent of $t_{\mathrm{max}}$)
for the on-transition case \textbf{(iii)} with
$\omega_1 = 4\pi$ and $t_{\mathrm{max}} = \HALF$ sec.
The solid lines are the simulated results for these
components, while the dashed line depicts the
evolution of the diagonal $\IZ^\spinA$
(starting from $1$) and $\IZ^\spinA
\IZ^\spinB$ (starting from $0$) components.
The symbols in the plot are the same
as in Figure \ref{fig:seriesA}.
} \label{fig:seriesC} \end{figure}

In order to demonstrate that these effective Hamiltonians
correctly describe the evolution of a weakly-coupled
two-spin system under a monochromatic RF field,
we used a solution of triply-labeled ${}^{13}\mathsf{C}$ alanine
($\mathsf{CO}_2^--\mathsf{CH}[-\mathsf{NH}_3^+]-\mathsf{CH}_3$)
in $\mathsf{D}_2\mathsf{O}$.
The carboxyl carbon was treated as the ``$\spinA$'' (target)
spin, the alpha-carbon as the ``$\spinB$'' (control) spin.
Since the carboxyl to methyl carbon coupling
constant of ca.~$1.4$ Hz.~is less than the
peak width, its effects could be ignored, and
the effects of the protons eliminated by decoupling.
The following three experiments were performed at $16$
evenly spaced time points in the interval from
$t_{\mathrm{max}}/16$ to $t_{\mathrm{max}}$:
\renewcommand{\theenumi}{\roman{enumi}}
\renewcommand{\labelenumi}{(\textbf\theenumi)}
\begin{enumerate}
\item On the $\omega^\spinA$ resonance
with $\omega_1 = \pi |J^{\spinA\spinB}|$
($t_{\mathrm{max}} = \sqrt{2}/|J^{\spinA\spinB}|$).
\item On the $\omega^\spinA + \pi J^{\spinA\spinB}$
transition with $\omega_1 = 2\pi |J^{\spinA\spinB}|$
($t_{\mathrm{max}} = 1/|J^{\spinA\spinB}|$).
\item On the $\omega^\spinA + \pi J^{\spinA\spinB}$
transition with $\omega_1 = 4 \pi = 2 \pi \times 2$
Hz.~($t_{\mathrm{max}} = \pi/\omega_1 = \HALF$ sec.).
\end{enumerate}
Experiments \textbf{(i)} and \textbf{(iii)} validate
the correctness of the effective Hamiltonian in the
two cases that may be used to implement a c-NOT gate,
while \textbf{(ii)} validates the theory for case in
which the evolution of coherence is more complicated.
All the experiments were performed with the spin system
initially at equilibrium in a $9.4$ Tesla field.

After the RF field had been applied
for each of the $16$ time periods used,
a 24k point FID (free-induction decay) was
collected over $590$ msec., zero-filled to 48k,
and Fourier transformed to yield the complete spectrum.
A 128 point window centered on the peaks from the $\spinA$-spin
was taken and the absorptive in-phase ($\IX^\spinA$),
absorptive anti-phase ($2 \IX^\spinA \IZ^\spinB$),
dispersive anti-phase ($2 \IY^\spinA \IZ^\spinB$)
and dispersive in-phase ($\IY^\spinA$)
components of the density matrix extracted from it.
This was done by a least-squares fit of a linear
combination of model peak shapes for each of
these four components to the spectrum window,
where the model peak shapes were computed using
a coupling constant of $54$ Hz. together with
a Lorentzian peak shape of half-width $0.85$ Hz.

In order to compare these results with the theory,
the results predicted from the above effective
Hamiltonians were also computed by numerical simulation,
using the same coupling constant and resonance frequencies.
These results were normalized so that the initial
$\IZ^\spinA$ state had unit norm,
and the experimental fits scaled so that the
root-mean-square values of the experimental
results were the same as the simulated results.
If necessary, the spectrum was adjusted by means
of a first-order phase correction so as to obtain
the best possible fit, as judged visually.
The simulations and final fits to the data for
each of these three series of experiments are shown
in Figures \ref{fig:seriesA} -- \ref{fig:seriesC}.
It should be clearly understood that the simulated curves
were {\em not\/} fitted to the experiments save by scaling.

Given the many sources of systematic error
present in NMR spectroscopy \cite{HochStern:96},
and the fact that the fits were not systematically
optimized with respect to the nonlinear parameters
(i.e.\ the coupling constant, resonance frequencies,
peak widths/shapes and spectrum phases),
the match of the experimental to the simulated results
is strong evidence for the validity of the theory.
The corresponding simulations using the transition
Hamiltonian (not shown) are essentially the same,
except that the high-frequency oscillations are not present.

\section{DERIVATION OF THE CONDITIONAL PHASE FACTORS}

\subsection{The on-resonance case}

Under the assumption of weak coupling,
it is shown in {\bf Appendix B} that if
$\omega_1 \ll |\omega^\spinA-\omega^\spinB|$,
the off-resonance effects due to the
$\IX^\spinB$ term in $\Heff$ can be ignored.
This can be understood intuitively through
the effective fields picture for the $\spinB$,
since under these fields are essentially along
the $z$-axis and hence only induce phase shifts.
Thus let $\Heff^0 \equiv \Heff - \omega_1 \IX^\spinB$,
\begin{equation} \label{eq:c1c2def}
c_1 ~\equiv~ \pi J^{\spinA\spinB} / \omega_1 \quad\text{and}\quad
c_2 ~\equiv~ (\omega_0^\spinA - \omega_0^\spinB) / \omega_1 ~,
\end{equation}
so that
\begin{equation} \begin{split}
\Heff^0 / \omega_1
~=~ & c_2 \IZ^\spinB +
( \IX^\spinA + 2 c_1 \VEC
I_\axisZ^\spinA \IZ^\spinB )
( \EM^\spinB + \EP^\spinB ) \\
=~ & c_2 \IZ^\spinB +
( \IX^\spinA - c_1 \IZ^\spinA )
\EM^\spinB +
( \IX^\spinA + c_1 \IZ^\spinA )
\EP^\spinB ~.
\end{split} \end{equation}
Since all three terms in this expression commute,
it follows that $\exp(-\UPS t \Heff^0 )$
\begin{equation}
=~ e^{-\UPS \omega_1 t \, (\IX^\spinA -
c_1 \IZ^\spinA) \EM^\spinB}
e^{-\UPS \omega_1 t \, (\IX^\spinA +
c_1 \IZ^\spinA) \EP^\spinB}
e^{-\UPS \omega_1 t c_2 \IZ^\spinB} ,
\end{equation}
where by the formula for the exponential of an operator
multiplied by a commuting idempotent \cite{SomCorHav:98},
$\exp(-\UPS \omega_1 t \, (\IX^\spinA \pm
c_1 \IZ^\spinA) \VEC E_{\pm}^\spinB)$
\begin{equation}
~=~ \VEC E_{\mp}^\spinB + \left(
\cos( \HALF\omega_1t{\scriptstyle\sqrt{1 + c_1^2}} )
- \frac{2\UPS (\IX^\spinA \pm c_1 \IZ^\spinA)}
{\scriptstyle\sqrt{1 + c_1^2}}
\sin( \HALF\omega_1t{\scriptstyle\sqrt{1 + c_1^2}} )
\right) \VEC E_{\pm}^\spinB ~.
\end{equation}
For $\omega_1 \equiv \pi |J^{\spinA\spinB}|$
and $t \equiv \pi / (\sqrt{2} \omega_1)$,
this implies $\exp(-\UPS t \Heff^0)$
\begin{equation} \begin{split}
~=~ & -\UPS\sqrt{2} \left(
(\IX^\spinA - \IZ^\spinA)
\EM^\spinB +
(\IX^\spinA + \IZ^\spinA)
\EP^\spinB \right)
e^{-\UPS \pi c_2 \IZ^\spinB / \sqrt{2}} \\
=~ & -2\UPS \left( \IX^\spinA \EM^\spinB
+ \IZ^\spinA \EP^\spinB \right)
e^{\UPS \pi \IY^\spinA / 2}
e^{-\UPS \pi c_2 \IZ^\spinB / \sqrt{2}} \\
=~ & e^{-\UPS \pi \IY^\spinA / 2} \,
e^{-\UPS \pi \IX^\spinA \EM^\spinB} \,
e^{-\UPS \pi \IZ^\spinA \EP^\spinB} \,
e^{-\UPS \pi c_2 \IZ^\spinB / \sqrt{2}} ~,
\end{split} \end{equation}
where have we used the fact that $\IY^\spinA$
anticommutes with $\IX^\spinA$ and $\IZ^\spinA$
to move its exponential to the far left.
Thus the applications of an on-resonance
pulse with $\omega_1 = \pi |J^{\spinA\spinB}|$
and $t = \pi / (\sqrt{2} \omega_1)$, followed by
an ordinary soft $-\pi/2$ $y$-pulse on spin $\spinA$,
yields the c-NOT gate up to a conditional phase
prefactor of $\exp(-\UPS \pi (\IZ^\spinA / 2 +
(c_2 \sqrt{2} - 1) \IZ^\spinB / 2 + \IZ^\spinA \IZ^\spinB))$.
By moving $\exp(\UPS\pi\IY^\spinA/2)$ to the far right instead,
one sees that one could just as well precede the on-resonance
pulse with a soft $+\pi/2$ $y$-pulse on spin $\spinA$.

\subsection{The on-transition case}
We begin again by breaking $\Heff^0
\equiv \Heff - \omega_1 \IX^\spinB$
into three commuting parts as above,
namely
\begin{equation} \begin{split} \label{eq:split}
(\pi/\omega_1) \Heff^0
~=~ & \pi \left( (c_1 + c_2) \IZ^\spinB
+ 2 c_1 \IZ^\spinA \EP^\spinB
+ \IX^\spinA (\EP^\spinB
+ \EM^\spinB) \right) \\
=~ & \pi \left( (c_1 + c_2) \IZ^\spinB
+ (2 c_1 \IZ^\spinA +
\IX^\spinA) \EP^\spinB
+ \IX^\spinA \EM^\spinB \right) \\
=~ & \VEC P + \VEC Q_\plus \EP^\spinB
+ \VEC Q_\minus \EM^\spinB ~,
\end{split} \end{equation}
where
\begin{equation}
\VEC P ~\equiv~ \pi (c_1 + c_2) \IZ^\spinB ,\quad
\VEC Q_\plus ~\equiv~ \pi (2 c_1 \IZ^\spinA
+ \IX^\spinA) ,\quad\text{and}\quad
\VEC Q_\minus ~\equiv~ \pi \IX^\spinA ~.
\end{equation}
Note that $\VEC P$ is diagonal (i.e.\ along the $z$-axis)
and that $\Htrn \equiv \VEC Q_\minus\EM^\spinB$,
but since $\VEC Q_\plus\EP^\spinB$ is not diagonal, 
$\exp(-\UPS (\pi/\omega_1) \Heff^0)$ and
$\exp(-\UPS \Htrn)$ do not simply differ by
conditional phases as in the on-resonance case.
Nevertheless, $\VEC Q_\plus\EP^\spinB$
can be readily diagonalized, because
\begin{equation} \label{eq:qplus}
\VEC Q_{\plus} ~=~ \pi (e^{-\UPS \theta \IY^\spinA} \IZ^\spinA
e^{\UPS \theta \IY^\spinA}) {\textstyle\sqrt{1 + 4 c_1^2}} ~,
\end{equation}
where $\theta$ is given by $\arctan(1/(2c_1))$.

To show that the effect of this diagonalization
upon the propagator $\exp(-\UPS t \Heff^0)$ is small,
we evaluate the norm of the difference of the
propagator with and without this transformation.%
\footnote{The norm we use is the square-root of the
geometric algebra ``scalar part'' (denoted by angular
brackets $\langle\cdot\rangle$) \cite{SomCorHav:98} of
the product of the quantity with its Hermitian conjugate.
For the two spin system considered here the scalar part is four
times the trace in the usual product Pauli matrix representation,
and hence our norm is just twice the standard Frobenius norm.}
Since $\IY^\spinA$ and $\VEC Q_\minus = \pi\IX^\spinA$ don't commute,
we operate with $\exp(-\UPS\theta\IY^\spinA\EP^\spinB)$
instead of $\exp(-\UPS\theta\IY^\spinA)$, obtaining
\begin{equation} \begin{split} \label{eq:horror}
& \left\| e^{-\UPS (\pi/\omega_1) \Heff^0}
- e^{\UPS \theta \IY^\spinA\EP^\spinB}
e^{-\UPS (\pi/\omega_1) \Heff^0}
e^{-\UPS \theta \IY^\spinA\EP^\spinB} \right\|^2 \\
=~ & \left\| e^{-\UPS \VEC Q_\plus\EP^\spinB}
- e^{\UPS \theta \IY^\spinA}
e^{-\UPS \VEC Q_\plus\EP^\spinB}
e^{-\UPS \theta \IY^\spinA} \right\|^2 \\
=~ & 2 - 2 \left\langle
e^{-\UPS \pi (2\IZ^\spinA+\IX^\spinA)\EP^\spinB}
e^{\UPS \pi \sqrt{1+4c_1^2} \IZ^\spinA \EP^\spinB}
\right\rangle \\
=~ & 2 - 2 \left\langle
\left( \EM^\spinB + \EP^\spinB
e^{-\UPS \pi (2\IZ^\spinA+\IX^\spinA)} \right)
\left( \EM^\spinB + \EP^\spinB
e^{\UPS \pi \sqrt{1+4c_1^2} \IZ^\spinA} \right)
\right\rangle \\
=~ & 1 - \left\langle
\left( \cos(\tfrac{\pi}{2}{\scriptstyle\sqrt{1+4c_1^2}}) - \UPS
\frac{2c_1\IZ^\spinA+\IX^\spinA}{\SHALF\scriptstyle\sqrt{1+4c_1^2}}
\sin(\tfrac{\pi}{2}{\scriptstyle\sqrt{1+4c_1^2}}) \right)
\right. \\ & \quad\quad\quad \left. \left( \rule[0pt]{0pt}{16pt}
\cos(\tfrac{\pi}{2}{\scriptstyle\sqrt{1+4c_1^2}}) + \UPS 2
\IZ^\spinA \sin(\tfrac{\pi}{2}{\scriptstyle\sqrt{1+4c_1^2}}) \right)
\right\rangle \\
=~ & 1 - \cos^2(\tfrac{\pi}{2}{\scriptstyle\sqrt{1+4c_1^2}})
- \sin^2(\tfrac{\pi}{2}{\scriptstyle\sqrt{1+4c_1^2}})
\frac{2c_1}{\scriptstyle\sqrt{1+4c_1^2}} \\
=~ & \sin^2(\tfrac{\pi}{2}{\scriptstyle\sqrt{1+4c_1^2}})
\left( 1 - 2c_1 / {\scriptstyle\sqrt{1+4c_1^2}} \right)
\end{split} \end{equation}

The second factor can be converted into the simple bound
\begin{equation}
\frac{1 + 4c_1^2 - 2 |c_1| {\scriptstyle\sqrt{1 + 4c_1^2}}}{1 + 4c_1^2} \le~
\frac{1}{4c_1^2} ~=~ \frac{\omega_1^2}{4(\pi J^{\spinA\spinB})^2} ~\ll~ 1 ~,
\end{equation}
thus showing that
$\exp(\UPS\VEC Q_\plus\EP^\spinB)$ is essentially a diagonal matrix
$\exp(\UPS\pi{\scriptstyle\sqrt{1+4c_1^2}}\IZ^\spinA\EP^\spinB)$
of phase factors as long as $\omega_1 \ll \pi|J^{\spinA\spinB}|$.
Noting that $\IZ^\spinA\EP^\spinB$ still commutes with $\Htrn$,
it follows that
\begin{equation}
e^{-\UPS(\pi/\omega_1)\Heff^0} ~\approx~
e^{-\UPS\Htrn} e^{-\UPS \pi(c_1+c_2) \IZ^\spinB}
e^{-\UPS \pi\sqrt{1+4c_1^2} \IZ^\spinA\EP^\spinB} ~,
\end{equation}
where the last two factors are conditional phases.

It is interesting to observe that since
$\sin^2(\frac{\pi}{2} {\scriptstyle\sqrt{1+4c_1^2}}) = 0$
if ${\textstyle\sqrt{1+4c_1^2}} = 2n$ for an integer $n > 0$,
an exact implementation of the c-NOT is obtained when
$|c_1| \equiv |\pi J^{\spinA\spinB}|/\omega_1 = \HALF\sqrt{4n^2-1}$
or $\omega_1 = 2\pi|J^{\spinA\spinB}| / \sqrt{4n^2-1}$,
in which case the time required for the c-NOT is
$\pi/\omega_1 = \sqrt{4n^2-1} / |2J^{\spinA\spinB}|$.
For $n = 1$, this is $\sqrt{3}/|2J^{\spinA\spinB}|$,
which is longer than the $1/|\sqrt{2}J^{\spinA\spinB}|$
required by the on-resonance implementation,
although in some circumstances the fact that
an additional soft pulse is not needed in
the on-transition case might be an advantage.

\section{CONCLUSIONS}
We have presented a detailed analysis of two
Pound-Overhauser implementations of the controlled-NOT gate,
one using an on-transition $\axisX$-pulse
with $\omega_1 \ll 2\pi |J^{\spinA\spinB}|$,
and the other using an on-resonance $\axisX$-pulse
of power $\omega_1 = \pi|J^{\spinA\spinB}|$
and duration $1/(\sqrt{2}|J^{\spinA\spinB}|)$
followed by a soft $\pi \IY^\spinA$ pulse.
The correctness of the effective Hamiltonians for both
implementations have been validated by NMR experiments,
and the phase corrections derived using geometric algebra.
In the course of this analysis, it was shown that
the time required for the on-transition implementation
could be decreased to $t = \sqrt{3} / (2|J^{\spinA\spinB}|)$
while actually making the implementation exact (up to conditional phases)
by increasing the power to $\omega_1 = 2\pi|J^{\spinA\spinB}|/\sqrt{3}$.
Neither the on-resonance nor the on-transition
Pound-Overhauser implementation is as efficient
as the pulse sequence in \cite{CorPriHav:98}, which
takes only $1/(2|J^{\spinA\spinB}|)$ plus two soft pulses.
Nevertheless, the Pound-Overhauser implementations have
the advantage that in systems of more than two spins,
it should be easier to find a sequence of $\pi$
pulses which refocuses the evolution of all the
remaining spins while the gate is in progress,
because it is not necessary to treat the spins
to which the gate is being applied specially.
Geometric algebra calculations with sums of transition
Hamiltonians \cite{SomCorHav:98} further indicate that
the simultaneous irradiation of multiple resonances,
which we call \emph{compound pulses}, can be used
to directly excite multiple quantum transitions,
and thereby accomplish in a single step what
might otherwise take considerably more time.
Their effective Hamiltonians will
be the subject of a future paper.

\ack
This work was supported by the U.S. Army Research
Office under grant number DAAG 55-97-1-0342
from the DARPA Ultrascale Computing Program.

\appendix
\section{DIAGONALIZATION OF THE ON-RESONANCE HAMILTONIAN}
In the on-resonance case, the characteristic
polynomial $\det( \Heff / \omega_1 - \lambda )$ of
the effective Hamiltonian is a quadratic in $\lambda^2$,
with eigenvalues
\begin{equation}
\lambda_{\pm}^2 ~=~ \tfrac{1}{4} \left( 2 + c_1^2 + c_2^2
\pm 2 {\textstyle\sqrt{1 + c_2^2(1 + c_1^2)}} \right) ~,
\end{equation}
where $c_1$ and $c_2$ are given by Eq.\ (\ref{eq:c1c2def}).
It is therefore reasonable to expect that it will be
possible to analytically diagonalize this Hamiltonian,
and we now show how this can be done using
geometric algebra \cite{SomCorHav:98}.

We begin by rewriting the on-resonance Hamilonian
(Eq.\ (\ref{eq:onres})) in the form
\begin{equation}
\Heff / \omega_1 ~=~
2 c_1 \IZ^\spinA \IZ^\spinB
+ \IX^\spinA + c_2' \IZ^\spinB
\exp(\UPS \mu 2 \IY^\spinB) ~,
\end{equation}
where
\begin{equation} \label{eq:c2prime}
c_2' ~=~ {\textstyle\sqrt{1 + c_2^2}} \quad\text{and}\quad
\tan(\mu) ~=~ 1 / c_2 ~.
\end{equation}
Rotating spin $\mu$ about $2\IY^\spinB$ this yields
\begin{equation} \begin{split}
\Heff' / \omega_1
~\equiv~ & e^{\UPS \mu \IY^\spinB}
(\Heff / \omega_1) e^{-\UPS \mu \IY^\spinB} \\
=~& 2 c_1 \IZ^\spinA \exp(-\UPS \mu 2 \IY^\spinB)
\IZ^\spinB + \IX^\spinA + c_2' \IZ^\spinB \\
=~& 2 c_1 \IZ^\spinA \left( \IZ^\spinB
\cos(\mu) + \IX^\spinB \sin(\mu) \right)
+ \IX^\spinA + c_2' \IZ^\spinB \\
=~& e^{-\UPS \nu 2 \IY^\spinA \IZ^\spinB}
\left( 2 c_1' \IZ^\spinA \IZ^\spinB
+ 2 c_1 \IZ^\spinA \IX^\spinB \sin(\mu)
+ c_2' \IZ^\spinB \right) e^{\UPS \nu 2
\IY^\spinA \IZ^\spinB} ~,
\end{split} \end{equation}
where
\begin{equation}
c_1' ~=~ {\textstyle\sqrt{1 + c_1^2 \cos^2(\mu)}}
\quad\text{and}\quad
\tan(\nu) ~=~ 1 / (c_1 \cos(\mu)) ~.
\end{equation}
The outer exponentials can be eliminated
by applying the opposite rotation, obtaining
\begin{equation} \begin{split} \label{eq:Hdprime}
\Heff'' / \omega_1
~\equiv~ & e^{\UPS \nu 2 \IY^\spinA
\IZ^\spinB} (\Heff' / \omega_1)
e^{-\UPS \nu 2 \IY^\spinA \IZ^\spinB} \\
=~& 2 c_1' \IZ^\spinA \IZ^\spinB
+ 2 c_1 \IZ^\spinA \IX^\spinB
\sin(\mu) + c_2' \IZ^\spinB ~.
\end{split} \end{equation}

To complete the diagonalization, this is rewritten as
\begin{equation} \begin{split}
\Heff'' / \omega_1
~=~ & \left( (c_1' + c_2') \IZ^\spinB
+ c_1 \sin(\mu) \IX^\spinB \right) \EP^\spinA \,- \\
& \left( (c_1' - c_2') \IZ^\spinB
+ c_1 \sin(\mu) \IX^\spinB \right) \EM^\spinA \\
=~& 2 \lambda_\plus \IZ^\spinB
e^{-\UPS \kappa_\plus 2 \IY^\spinB} \EP^\spinA
- 2 \lambda_\minus \IZ^\spinB
e^{-\UPS \kappa_\minus 2 \IY^\spinB} \EM^\spinA ~,
\end{split} \end{equation}
where
\begin{equation}
\lambda_\pm ~=~ \HALF \sqrt{(c_1' \pm c_2')^2 + (c_1 \sin(\mu))^2}
\end{equation}
are the eigenvalues as above, and
\begin{equation}
\tan(\kappa_\pm) ~=~ \frac{c_1\sin(\mu)}{c_1' \pm c_2'} ~.
\end{equation}
Therefore the conditional rotation
\begin{equation} \begin{split}
\VEC K ~\equiv~ & e^{-\UPS \kappa_\plus \IY^\spinB}
\EP^\spinA + e^{-\UPS \kappa_\minus \IY^\spinB}
\EM^\spinA \\
=~& e^{-\UPS \kappa_\plus \EP^\spinA \IY^\spinB}
e^{-\UPS \kappa_\minus \EM^\spinA \IY^\spinB}
\end{split} \end{equation}
completes the diagonalization to
\begin{equation} \begin{split}
\VEC K (\Heff'' / \omega_1) \tilde{\VEC K} ~=~
& 2 \lambda_\plus \EP^\spinA \IZ^\spinB -
2 \lambda_\minus \EM^\spinA \IZ^\spinB \\
=~& \lambda_\plus \EP^\spinA \EP^\spinB
- \lambda_\plus \EP^\spinA \EM^\spinB
- \lambda_\minus \EM^\spinA \EP^\spinB
+ \lambda_\minus \EM^\spinA \EM^\spinB ~.
\end{split} \end{equation}

While it is possible to write down
the time-dependent exponential of
the diagonalized Hamiltonian, and
to transform it back to that of
the original Hamiltonian $\Heff$,
the resulting propagator is too complicated
to yield much insight into the dynamics.
Nevertheless, in the next Appendix
the above diagonal form will enable us to
give an elementary proof that off-resonance
effects can be neglected in the on-resonance case.

\section{OFF-RESONANCE EFFECTS ON THE PROPAGATORS}

\subsection{The on-resonance case}
In order to justify the above assumption
that the $\IX^\spinB$ term
in the Hamiltonian can be neglected,
let us extend our definition of $\Heff$
(Eq.~\ref{eq:onres})  to
\begin{equation}
\Heff^\alpha ~\equiv~ 2 \pi J^{\spinA\spinB}
\IZ^\spinA \IZ^\spinB
+ (\omega_0^\spinA - \omega_0^\spinB) \IZ^\spinB
+ \omega_1 (\IX^\spinA + \alpha \IX^\spinB) ~.
\end{equation}
Note that $\left.\Heff^\alpha\right|_{\alpha=0} = \Heff^0
\equiv \Heff - \omega_1 \IX^\spinB$ as above.
Accordingly, Eq.~(\ref{eq:c2prime}) is replaced by
\begin{equation}
c_2' ~=~ {\textstyle\sqrt{\alpha^2 + c_2^2}}
\quad\text{and}\quad \tan(\mu) ~=~ \alpha / c_2 ~.
\end{equation}
The rest of the diagonalization of $\Heff^\alpha$
goes through exactly as for $\Heff = \Heff^1$.
It now follows from Eq.~(\ref{eq:Hdprime})
that $\exp(-\UPS t \Heff^\alpha)$
\begin{equation} \begin{split}
=~ & e^{-\UPS\mu\IY^\spinB}
e^{-\UPS\nu\IY^\spinA\IZ^\spinB}
e^{-\UPS\kappa_\plus\EP^\spinA\IY^\spinB}
e^{-\UPS\kappa_\minus\EM^\spinA\IY^\spinB}
e^{-\omega_1t\UPS\lambda_\plus\EP^\spinA\IZ^\spinB}
\\ & \qquad \times\,
e^{\omega_1t\UPS\lambda_\minus\EM^\spinA\IZ^\spinB}
e^{\UPS\kappa_\minus\EM^\spinA\IY^\spinB}
e^{\UPS\kappa_\plus\EP^\spinA\IY^\spinB}
e^{\UPS\nu\IY^\spinA\IZ^\spinB}
e^{\UPS\mu\IY^\spinB} ~. \label{eq:buggers}
\end{split} \end{equation}
where $\mu$, $\nu$, $\kappa_{\pm}$ and
$\lambda_{\pm}$ are all functions of $\alpha$.

We wish to show that the propagators
$\exp(-\UPS t \Heff^0)$ and $\exp(-\UPS t \Heff)$
are approximately equal.
Although this is implied by first-order perturbation theory,
for any noninfinitesimal perturbation this standard
argument falls short of being a rigorous proof.
To this end, let us fix $t$ and define the function
\footnote{
Since multiplying a propagator by a scalar phase factor
has no effect on how it transforms the density matrix,
the correct way to compare two $N\times N$ propagators
$\VEC U_1$, $\VEC U_2$ is to compute the expression
\begin{equation*} \begin{split}
1 - |\TR(\VEC U_1 \tilde{\VEC U}_2)/N|^2
~=~ & 1 - \AVG{\VEC U_1 \tilde{\VEC U}_2}^2
- \AVG{\UPS \VEC U_1 \tilde{\VEC U}_2}^2 \\
~=~ & \tfrac{1}{4} \| \VEC U_1 - \VEC U_2 \|^2
\| \VEC U_1 + \VEC U_2 \|^2 +
\tfrac{1}{4} \| \VEC U_1 - \UPS \VEC U_2 \|^2
\| \VEC U_1 + \UPS \VEC U_2 \|^2 - 1 ~.
\end{split} \end{equation*}
It is sufficient, however, to show
that just one of the above norms vanishes,
e.g.\ $\| \VEC U_1 - \VEC U_2 \|^2$,
meaning that the propagators are equal
without any overall phase differences,
and this turns out to be the case for all
the propagators considered in this paper.
}
\begin{equation} \begin{split}
f(\alpha) ~\equiv~ &
\HALF \left\| e^{-\UPS t \Heff^\alpha}
- e^{-\UPS t \Heff^0} \right\|^2 \\
=~ & 1 - \left\langle e^{-\UPS t \Heff^\alpha}
e^{\UPS t \Heff^0} \right\rangle ~.
\end{split} \end{equation}
Using the fact that $\| \VEC U \| = 1$ for unitary $\VEC U$,
the Cauchy-Schwarz and triangle inequalities, and the
invariance of the scalar part under cyclic permutations,
the magnitude of the derivative may be bounded as follows:
\begin{equation} \begin{split}
|f'(\alpha)| ~=~ & \left| \left\langle
\partial / \partial\alpha e^{-\UPS t \Heff^\alpha}
e^{\UPS t \Heff^0} \right\rangle \right| \\ \ms
\le~ & \left\| \partial/\partial\alpha
e^{-\UPS t \Heff^\alpha} \right\| \left\|	
\rule[0pt]{0pt}{10pt} \smash{e^{\UPS t \Heff^0}} \right\|
~=~ \left\| \partial/\partial\alpha
e^{-\UPS t \Heff^\alpha} \right\| \\ \ms
\le~ & 2 \| \UPS \IY^\spinB
\, \partial \mu / \partial \alpha \|
+ 2 \| \UPS 2 \IY^\spinA \IZ^\spinB
\, \partial \nu / \partial \alpha \| \\ & \quad
+\, 2 \| \UPS \EP^\spinA \IY^\spinB
\, \partial \kappa_\plus / \partial \alpha \|
+ 2 \| \UPS \EM^\spinA \IY^\spinB
\, \partial \kappa_\minus / \partial \alpha \| \\ & \quad
+\, \| \,\omega_1 t\, \UPS \EP^\spinA \IZ^\spinB \,
\partial \lambda_\plus / \partial \alpha \|
+ \| \,\omega_1 t\, \UPS \EM^\spinA \IZ^\spinB \,
\partial \lambda_\minus / \partial \alpha \|
\end{split} \end{equation}

To bound the first two terms on the right-hand side,
the derivatives are evaluated and simplified by means
of the assumptions $2 \le 1 + c_1^2 \le c_2^2$, i.e.
\begin{equation} \begin{split}
2\, \left\| \UPS \IY^\spinB \,
\partial \mu / \partial \alpha \right\|
~=~ & \left| \partial/\partial\alpha\, \arctan(\alpha/c_2) \right| \\
=~ & \frac{1}{|c_2|(1+(\alpha/c_2)^2)} ~\le~ \frac{1}{|c_2|} ~,
\end{split} \end{equation}
and (using the fact that
$\cos(\mu) = (1 + (\alpha/c_2)^2)^{-1/2}$)
\begin{equation} \begin{split}
2 \left\| \UPS 2 \IY^\spinA \IZ^\spinB
\, \partial \nu / \partial \alpha \right\|
~=~ & \, \left| \partial / \partial \alpha
\arctan(1/(c_1\cos(\mu))) \right| \\
=~ & \, \left| \partial / \partial \alpha \arctan({\textstyle
\sqrt{\alpha^2 + c_2^2}} / (c_1c_2)) \right| \\
=~ & \, \frac{\alpha|c_1||c_2|}{(\alpha^2 + c_2^2(1 + c_1^2))
\sqrt{\alpha^2+c_2^2}} \\
\le~ & \frac{|c_1|}{(1+c_1^2)c_2^2} ~\le~ \frac{1}{|c_1|c_2^2} ~,
\end{split} \end{equation}
where the inequality is obtained by setting $\alpha = 1$
in the numerator and $\alpha = 0$ in the denominator.

To bound the terms depending on $\kappa_\pm$,
define $\xi \equiv \alpha^2 + c_2^2(1 + c_1^2)$,
and proceed as follows:
\begin{equation} \begin{split}
2 \left\| \UPS \VEC E_\pm^\spinA \IY^\spinB
\, \partial \kappa_\pm / \partial \alpha \right\|
~=~ & 2^{-1/2} \left| \partial / \partial \alpha \,
\arctan\left( c_1\sin(\mu) / (c_1' \pm c_2') \right) \right| \\
=~ & \frac1{\sqrt2} \left| \frac{\partial}{\partial \alpha}
\arctan\left( \frac{\alpha c_1}{\sqrt{\xi}
\pm (\alpha^2 + c_2^2)} \right) \right| \\
=~ & \frac1{\sqrt2} \left| \frac{c_1 ((c_2^2 - \alpha^2)
\sqrt{\xi} \pm c_2^2 (1 + c_1^2))}
{(\alpha^2+c_2^2)((1+\alpha^2+c_1^2+c_2^2)
\sqrt{\xi} \pm 2 \xi)} \right|
\end{split} \end{equation}
An elementary analysis of the denominator
shows that it is nonnegative and reaches its
minimum in the interval $[0,1]$ at $\alpha = 0$
in both the ``$+$'' and ``$-$'' cases,
while the numerator is likewise nonnegative
and reaches its maximum at $\alpha = 0$ in the
``$+$'' case and $\alpha = 1$ in the ``$-$''.
This together with further simplifications
lead to the bounds
\begin{equation}
\frac1{\sqrt2} \left| \frac{\partial \kappa_\plus}{\partial \alpha} \right|
~\le~ \frac{|c_1|}{\sqrt2 c_2^2}
\qquad\text{and}\qquad
\frac1{\sqrt2} \left| \frac{\partial \kappa_\minus}{\partial \alpha} \right|
~\le~ \frac{|c_1|}{\sqrt2|c_2|(|c_2| - \sqrt2|c_1|)} ~.
\end{equation}

Lastly, the terms depending on $\lambda_\pm$ are
\begin{equation} \begin{split}
\left\| \pm \omega_1 t\, \UPS \VEC E_\pm^\spinA \IZ^\spinB
\, \partial \lambda_\pm / \partial \alpha \right\|
~=~ & \frac{\omega_1 t}{2\sqrt2} \left| \partial / \partial\alpha
\, \HALF \sqrt{(c_1' \pm c_2')^2 + (c_1 \sin(\mu))^2} \right| \\
=~ & \frac{\omega_1 t}{4\sqrt2} \left| \frac{\alpha (\sqrt{\xi} \pm 1)}
{\sqrt{\xi(1+\alpha^2+c_1^2+c_2^2 \pm 2\sqrt{\xi})}} \right| ~.
\end{split} \end{equation}
As before, an elementary analysis shows
that in both the ``$+$'' and ``$-$'' cases,
the denominator is minimized at $\alpha = 0$,
while the numerator is maximized at $\alpha = 1$.
This together with further simplifications leads to
\begin{equation}
\frac{\omega_1 t}{2\sqrt2} \left| \frac{\partial\lambda_\plus}{\partial\alpha}
\right| ~\le~ \frac{\omega_1 t}{4\sqrt2|c_2|}
\qquad\text{and}\qquad
\frac{\omega_1 t}{2\sqrt2} \left| \frac{\partial\lambda_\minus}{\partial\alpha}
\right| ~\le~ \frac{\omega_1 t}{4(|c_2|-\sqrt2|c_1|)} ~.
\end{equation}

Recalling that in the on-resonance case
$\omega_1 = \pi |J^{\spinA\spinB}|$,
so that $|c_1| = 1$ and $\omega_1t \le
\omega_1 t_{\mathrm{max}} = \pi/\sqrt2$,
we finally obtain
\begin{equation}
|f'(\alpha)| ~\le~ \frac1{|c_2|} + \frac1{c_2^2}
+ \frac1{\sqrt2c_2^2} + \frac1{\sqrt2|c_2|(|c_2| - \sqrt2)}
+ \frac{\pi}{8|c_2|} + \frac{\pi/\sqrt2}{4(|c_2| - \sqrt2)}
\end{equation}
for all $0 \le \alpha \le 1$.
Letting $g(c_2)$ denote the right-hand side of this inequality,
it now follows from our assumption $|c_2| \equiv
|\omega_0^\spinA - \omega_0^\spinB| / \omega_1 \gg 1$ that
\begin{equation} \begin{split}
f(1) ~\equiv~ & \HALF \left\| e^{-\UPS t \Heff}
- e^{-\UPS t \Heff^0} \right\|^2 \\
\le~ & f(0) + \max_{0\le\alpha\le 1} |f'(\alpha)| \\
=~ & g(c_2) ~\ll~ 1 ~,
\end{split} \end{equation}
as desired.

\subsection{The on-transition case}
It remains to be shown that
$\exp(-\UPS (\pi/\omega_1) \Heff)
\approx \exp(-\UPS (\pi/\omega_1) \Heff^0)$
in the on-transition case.
This is again expected from first-order
perturbation theory, but a rigorous proof
is rendered nontrivial by the fact that
$\Heff$ admits no closed-form diagonalization.
The proof presented here is based upon the ``sinch'' commutator
series expansion of the directional derivative of the matrix
exponential which may be found in \cite{NajfeHavel:95},
using however geometric algebra to proceed in
a coordinate-free manner \cite{SomCorHav:98}.

Thus we define the perturbed Hamiltonian $\Heff^\alpha
\equiv \Heff^0 + \alpha\, \omega_1 \IX^\spinB$,
so that $\exp(-\UPS t \Heff^\alpha)$
\begin{equation} \label{eq:taylor}
=~ e^{-\UPS t \Heff^0} - \alpha\,
\UPS t \omega_1 \IX^\spinB \cdot \nabla
e^{-\UPS t \Heff^0} + O(\alpha^2) ~.
\end{equation}
The ``sinch'' expansion of the direction derivative is%
\footnote{The missing factor of ``$t$'', as compared
with Eq.\ (105) in \cite{NajfeHavel:95}, is due to
the fact that we are treating here the time multiplying
the Hamiltonian and the time multiplying the perturbation
as independent parameters, and dropping the latter.}
\begin{equation} \begin{split} \label{eq:sinch}
& \IX^\spinB \cdot \nabla
e^{-\UPS t \Heff^\alpha} ~\equiv~
\tfrac{\UPS}{t\omega_1} \tfrac{\mathrm d}{\mathrm d\alpha} 
e^{-\UPS t \Heff^\alpha} \\
=~& e^{-\UPS \SFRAC{t}{2} \Heff^\alpha}
\left( \sum_{k=0}^\infty \frac{\{ \IX^\spinB,
(\tfrac{t}{2} \Heff^\alpha)^{2k} \}}
{(-1)^k \, (2k+1)!} \right)
e^{-\UPS \SFRAC{t}{2} \Heff^\alpha} ~,
\end{split} \end{equation}
where the commutator powers are defined recursively by
\begin{equation} \begin{split}
& \{ \VEC X, \VEC Y^0 \} ~\equiv~ \VEC X ~,\quad\text{and} \\
& \{ \VEC X, \VEC Y^k \} ~\equiv~ [ \{ \VEC X, \VEC Y^{k-1} \}, \VEC Y ]
\quad (k > 0) ~.
\end{split} \end{equation}
The essential thing to note is that
$\{ \IX^\spinB, \tfrac{t}{2} \Heff^\alpha \}
= \{ \IX^\spinB, \tfrac{t}{2} \Heff^0 \}$,
so that the summation in Eq.\ (\ref{eq:sinch})
is independent of $\alpha$.

In order to evaluate the commutator powers explicitly,
one proceeds as follows:
\begin{align}
\{ \IX^\spinB, (\tfrac{t}{2} \Heff^0)^0 \}
~=~ & \IX^\spinB ~\text{by definition;} \nonumber \\
\rule[0pt]{0pt}{16pt} 
\{ \IX^\spinB, (\tfrac{t}{2} \Heff^0)^1 \}
~=~ & \omega_1 \tfrac{t}{2} [ \IX^\spinB,
(c_1 + c_2) \IZ^\spinB + 2c_1 \IZ^\spinA
\EP^\spinB + \IX^\spinA ] \nonumber \\
=~ & \omega_1 \tfrac{t}{2} ((c_1 + c_2) + 2c_1 \IZ^\spinA)
[ \IX^\spinB, \IZ^\spinB ] \nonumber \\
=~ & \omega_1 \tfrac{t}{2} ((c_1 + c_2) + 2c_1 \IZ^\spinA)
( -\tfrac{\UPS}{2} \IY^\spinB ) ~\text{;} \nonumber \\
\rule[0pt]{0pt}{16pt} 
\{ \IX^\spinB, (\tfrac{t}{2} \Heff^0)^2 \}
~=~ & \omega_1 \tfrac{t}{2} ((c_1 + c_2) + 2c_1 \IZ^\spinA)
[ -\tfrac{\UPS}{2} \IY^\spinB,
\tfrac{t}{2} \Heff^0 ] \nonumber \\
=~ & \left( \omega_1 \tfrac{t}{4} ((c_1 + c_2) + 2c_1 \IZ^\spinA)
\right)^2 \IX^\spinB ~\text{;} \nonumber \\
\intertext{and in general}
\{ \IX^\spinB, (\tfrac{t}{2} \Heff^0)^{2k} \}
~=~ & \left( \omega_1 \tfrac{t}{4} ((c_1 + c_2) + 2c_1 \IZ^\spinA)
\right)^{2k} \IX^\spinB ~\text{.} \nonumber
\end{align}
It follows that the commutator series is given by
\begin{equation} \begin{split}
\sum_{k=0}^\infty \frac{\{ \IX^\spinB,
(\tfrac{t}{2} \Heff^\alpha)^{2k} \}}{(-1)^k \, (2k+1)!}
~=~& \left( \sum_{k=0}^\infty \frac{
\left( \omega_1 \tfrac{t}{4} ((c_1 + c_2) + 2c_1 \IZ^\spinA)
\right)^{2k}}{(-1)^k \, (2k+1)!} \right) \IX^\spinB \\
=~& \FUN{sinc}(\omega_1 \tfrac{t}{4}
((c_1 + c_2) + 2c_1 \IZ^\spinA))
\IX^\spinB ~,
\end{split} \end{equation}
where $\FUN{sinc}(\VEC X) \equiv \sin(\VEC X)/\VEC X$ as usual.
This can be rewritten in terms of scalar functions as follows:
\begin{equation} \begin{split}
& \FUN{sinc}(\omega_1 \tfrac{t}{4}
((c_1 + c_2) + 2c_1 \IZ^\spinA))
(\EP^\spinA + \EM^\spinA) \\
=~& \FUN{sinc}(\omega_1 \tfrac{t}{4}
((c_1 + c_2) + c_1)) \EP^\spinA
+ \FUN{sinc}(\omega_1 \tfrac{t}{4}
((c_1 + c_2) - c_1)) \EM^\spinA ~.
\end{split} \end{equation}
Finally, on taking norms of both sides
of Eq.\ (\ref{eq:sinch}), one obtains
\begin{equation} \begin{split}
\| \IX^\spinB \cdot \nabla
e^{-\UPS t \Heff^\alpha} \|
~\le~& \| e^{-\UPS \SFRAC{t}{2} \Heff^\alpha} \|
\| \FUN{sinc}(\omega_1 \tfrac{t}{4} (c_2 + 2c_1))
\EP^\spinA \IX^\spinB \\
& +\, \FUN{sinc}(\omega_1 \tfrac{t}{4} c_2)
\EM^\spinA \IX^\spinB \|
\| e^{-\UPS \SFRAC{t}{2} \Heff^\alpha} \| \\
\le~& |\FUN{sinc}(\omega_1 \tfrac{t}{4} (c_1 + 2c_2))|
\| \EP^\spinA \IX^\spinB \| \\
& +\, |\FUN{sinc}(\omega_1 \tfrac{t}{4} c_2)|
\| \EM^\spinA \IX^\spinB \| \\
\le~& \tfrac{\sqrt2}{t\,\omega_1}
\left( |c_2 + 2c_1|^{-1} + |c_2|^{-1} \right) \\
\le~&  \frac{\sqrt2}{t\,\omega_1} \left(
\frac{2(|c_2|+|c_1|)}{|c_2|(|c_2|-2|c_1|)} \right) \\
\le~ & \sqrt8\left( t\,\omega_1(|c_2| - 2|c_1|) \right)^{-1} ~.
\end{split} \end{equation}

Now consider the function
\begin{equation} \begin{split}
f(\alpha) ~\equiv~ & \HALF
\left\| e^{-\UPS t \Heff^\alpha}
- e^{-\UPS t \Heff^0} \right\|^2 \\
=~& 1 - \left\langle e^{-\UPS t \Heff^\alpha}
e^{\UPS t \Heff^0} \right\rangle ~,
\end{split} \end{equation}
the magnitude of whose derivative may be bounded as follows:
\begin{equation} \begin{split}
\left| f'(\alpha) \right| ~=~ &
\left| \left\langle \UPS t \omega_1 \IX^\spinB
\cdot \nabla e^{-\UPS t \Heff^\alpha}
e^{\UPS t \Heff^0} \right\rangle \right| \\
\le~& t \omega_1 \| \IX^\spinB \cdot
\nabla e^{-\UPS t \Heff^\alpha} \|
\| e^{\UPS t \Heff^0} \| \\
\le~ & \sqrt8 \left( |c_2| - 2|c_1| \right)^{-1}
\end{split} \end{equation}
It follows that
\begin{equation} \begin{split}
f(1) ~\equiv~ & \HALF
\left\| e^{-\UPS t \Heff} -
e^{-\UPS t \Heff^0} \right\|^2 \\
\le~& f(0) + \max_{0\le\alpha\le 1} |f'(\alpha)| \\
\le~& \frac{\sqrt8\,\omega_1}{|\omega_0^\spinA - \omega_0^\spinB|
- 2 \pi |J^{\spinA\spinB}|} ~\ll~ 1
\end{split} \end{equation}
for all times $t$, as desired.

\bibliographystyle{jphysicsB}
\References
\bibliography{../../math,../../csci,../../nmr,%
../../phys,../../chem,../../self}
\endrefs

\end{document}